# Education Games To Learn Basic Algorithm With Near Isometric Projection Method

**Wirawan Istiono [1]**
Universitas Multimedia Nusantara
Scientia Boulevard Street, Gading Serpong, Curug Sangereng, Tangerang, Banten 15810, Indonesia

**Hijrah [2]**
Tengku Dirundeng Meulaboh Islamic State Institute (STAIN)
Sisingamangaraja Street, Gampa Meulaboh, Aceh 23116, Indonesia

**Nur Nawaningtyas.P [3]**
Institute of Information Management & Computer (STMIK) Widuri
Palmerah Barat Street, South Jakarta 12210, Indonesia

*Abstract* - **Basic programming and algorithm learning is one of the compulsory subjects required for students majoring in computers. As this lesson is knowledge base, it is very important and essential that before learn programmings languages students must be encourages to learn it to avoid difficulties that by using the algorithm learning games application with Near Isometric Projection, Students or prospective students become more interested in learning algorithms and programming. In this study, basic learning algorithms focused on the material Sequencing, Overloading, Procedures, Recursive Loops and Conditionals, which are made so that it can make it easier for students to learn the basics of programming algorithms**. The simulated results show that proposed Education Games with Near Isometric Projection method reach 83.87% statement of agreement that application games to learn basic programming algorithms were interesting and helped them to understand basic algorithm after testing using UAT. Testing with User Acceptance Test for 30 students of Multimedia Nusantara University

*Keyword-component; Education games, basic algorithm, basic programming, Isometric, Near Isometric Projection*.

## I. INTRODUCTION

Algorithm learning and programming are the basis for all computer majors, because this course is a basic subject before students plunge into case studies in programming languages. However, many students have difficulty in learning the subject of algorithm learning and programmings, this because the language structure is not as common as they use every day, and can also be caused by the following things according to Iain Milne, Glenn Rowe and Mark Guzdial[10][11], such as: wrong learning patterns, lack of quality references, lack of training and no interest.

In addition, for prospective students who intent to take major Subject as Department of Information, without knowing algorithms or programming languages there shall be obstacles for immediately understand the learning delivered in class. Alignment of algorithm subjects or not understanding students or prospective students about algorithms will certainly be a special obstacle for informatics students to be able to continue other programm languages. Due to the basis of all programming languages is an algorithm.

Games that use electronic media are interesting entertainment in the form of multimedia, many people young and old, prospective students or students are interested in games, so they can spend a lot of time playing games.

By trying to play the algorithm learning game application built using the Near Isometric Projection, researchist expect students or prospective students can easily understand and understand the flow of programming algorithms in a fun way, so that they can stimulate and increase students' enthusiasm to learn and understand algorithms and programming languages others.

Based on the identification and background described above, the problem can be formulated in this study are: How to design an algorithm learning games application can interest to users, both in terms of game play and user interface? and also how to determine the content of learning algorithm material, so that users can understand the learning of programming algorithms easily

The following is the limitations of the problem that will be applied in this study: The application of algorithm learning and programming games will be made using the Unity engine tools and C# programming language, algorithm learning and programming applications targeted at teenagers or adults and students or prospective Informatics students or other computer majors. Content The algorithm taught in this algorithm is a basic algorithm, which will be the basis of programming, which includes Sequencing, Overloading, Procedures, Recursive Loops and Conditionals. In this paper, it will be explained about the use of angle tiles in near isometric projection, and how to use a matrix to create levels, and also, will describe how characters move in the isometric world which are given looping and condition effects as element learning in these games. In the end this educational game was tested by students of Multimedia Nusantara University, to get a level of user acceptance of this educational game.

This paper is organized as follows. Section 2 contain related work of Algorithms and educational games, which are educational learning media, where with media can encourage students to think creatively and carry out activities with other students playing games in learning activities. Near Isometric Projection is defined in Section 3. Section 4 shows results of 83.87 % student agreed that this algorithm learning game was interesting and helped in understanding programming algorithms. Section 5 conclusion of paper and section 6 give reference to future work which could be done using adding more algorithms level with better level design.

## II. RELATED WORK

Algorithm is computational procedure that takes some value, or set of values, as input and produces some value, or set of values, as output or a sequence for calculation to solve a problem written in sequence [20]. So, programming algorithms are sequences or steps to solve computer programming problems. In programming, the important thing to understand is our logic in thinking about how to solve programming problems that will be made. For example, many math problems are easy if completed in writing, but quite difficult if we translate into programming. In this case, the algorithm and programming logic will be very important in solving problems.

Games as a learning media, which involves students in the process of experience and at the same time living the challenges, getting inspiration, motivated to think creatively, and integrate in activities with fellow students in playing games. Even though each student does the same activities with his friends, but the process of his inner experience in developing his own potential may vary. Games are facts that are analyzed to understand the process of behavior in a game, the choice of each decision in acting or saying becomes a conclusion as learning to produce oneself. Learning is the active process of students who develop their potential, while education is "a process, a method, an act of educating". [8]

Based on the explanation of the two words above, it can be concluded that the educational game, which is a learning media that is educational, where with the media can encourage students to think creatively and carry out activities with other students playing games in learning activities.

The term isometric comes from Greek, which means the same size (equal measure), in other words, isometric shows each side of the object with the same projection angle, which is 30 degrees from the horizontal line or the angle between the projection axes x, y, and z is 120 degrees. So that each side of the object gets the same portion in the field of work. Isometric used each side of the object can be seen (right, left, top), 2: 1 projection or commonly known as Near Isometric Projection has long been used since the era of pixel art, where the angle of use used is 26,565 degrees. This is because the 30 degree angle causes the object to look too steep, otherwise if the object is stored in pixel format, aliasing will occur (jagged display, curved on the line if displayed on a low resolution monitor). [17]

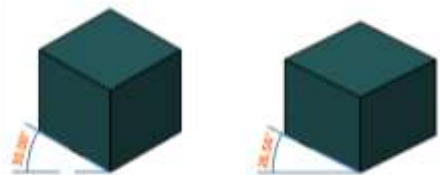

Figure 2.1. Tile with 30 degree angle and 26.56 degree

In isometrics, usually the size of the object's length, width, and height is calculated by unit grid. For example 1x1x2 means 1 grid length, 1 grid width and 2 grid height.

## III. METHOD

Isometric display is a method used to create 3D illusions for 2D games, this isometric method is commonly referred to as pseudo 3D or 2.5D. In making the initial isometric game used tile-base, each visual element is broken down into small pieces, called tiles, with a standard size. This tile will be arranged to form the game world according to predetermined data levels, usually determined by using 2D arrays. These tiles are each the same size, so that the tile length and tile width are the same [11].

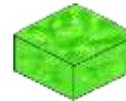

Figure 3.1. Sample Tile with near isometric projection

To create a level, 2-dimensional array will look like below, and the result from the array, will show like figure 3.2 below:

```
[[1,1,1,1,1],
 [1,0,1,0,1],
 [1,0,1,0,1],
 [1,1,1,1,1],
 [1,0,1,0,1],
 [1,0,1,0,1],
 [1,1,1,1,1]]
```

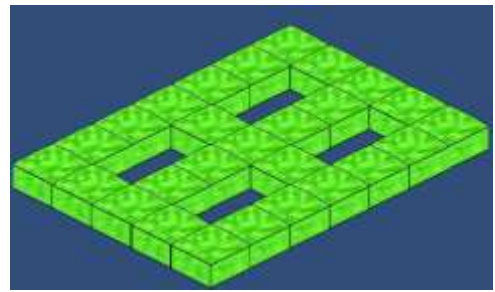

Figure 3.2. Result isometric array

Number 0 is to empty the tile, or not created, and number 1 is the command to make tiles with 1 level. To make a ladder or a higher level, the array can be written 2 for 2 levels, or 3 for level 3, as shown below.

```
[[1,1,1,1,1],
 [1,0,2,0,1],
 [1,0,2,0,1],
 [1,2,3,2,1],
 [1,0,2,0,1],
 [1,0,2,0,1],
 [1,1,1,1,1]]
```

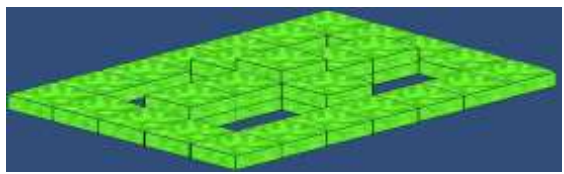

Figure 3.3. Result isometric array with levels

To make compilation, tiles can be arranged regularly with the right position like image above, the formula can be used as below :

positionX = (tile.sizeX / 2) – (nextTile.sizeX /2)
positionY = tile.sizeX / 4 + ((nextTile.sizeY – spaceHeightTile) / 2) + ((levelTile.sizeY – spaceHeightTile)/2)

because there are many tiles to make the game world, looping it's needed so the tiles can be arranged neatly and well. And this is sample code with C# language:

```csharp
levelMaker = new int[,] {
    {1,1,1,1,1},
    {1,0,2,0,1},
    {1,0,2,0,1},
    {1,2,3,2,1},
    {1,0,2,0,1},
    {1,0,2,0,1},
    {1,1,1,1,1}
};

int panjang = levelMaker.GetUpperBound(0);
int lebar = levelMaker.GetUpperBound(1);
float spaceHeightTile = 0.75f;

int orderSort = 0;
for (int i = 0; i <= panjang; i++) {
    for (int a = 0; a <= lebar; a++) {
        if (levelMaker[i, a] > 0) {

            for (int t = 0; t < levelMaker[i, a]; t++)
            {
                Vector2 pos = this.transform.position;
                pos.x = (a * (preFloor.GetComponent<SpriteRenderer>().bounds.size.x / 2)) - (i * preFloor.GetComponent<SpriteRenderer>().bounds.size.x / 2);
                pos.y = ((pos.y - (i * preFloor.GetComponent<SpriteRenderer>().bounds.size.x / 4)) + (pos.y - (a * ((preFloor.GetComponent<SpriteRenderer>().bounds.size.y - spaceHeightTile) / 2)))) + (t* ((preFloor.GetComponent<SpriteRenderer>().bounds.size.y - spaceHeightTile) / 2));
                GameObject objPre = Instantiate(preFloor, pos, Quaternion.identity) as GameObject;

                objPre.name = "tile" + a + i;
                orderSort++;
            }
        }
    }
}
```

To apply the character for education games, algorithm learning in isometric display, it needs to be created in accordance with the standard isometric by creating a character that has four-way or eight-way movements.

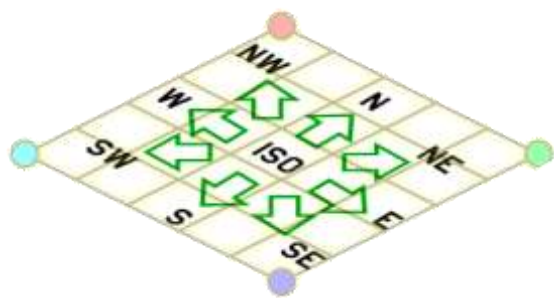

Figure 3.4. Eight-way character direction

Because one character movement must have eight directions, then if there is idle movement, walk or jump, it must be made 8 for each movement, which means that there must be at least 24 images needed for 3 movements, if only 1 image is needed for 1 movement 1 direction. The sample display of characters with 8 directions looks like the picture below.

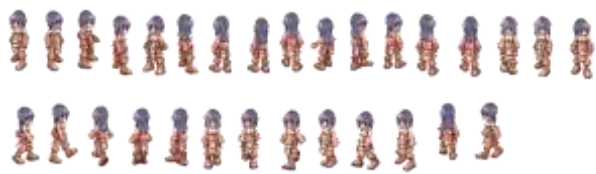

Figure 3.5. Sample character with 8 direction for every move

In addition to normal placement, handling accurate is needed to sort the depth to draw an isometric world. This ensures that items that are close to the player are drawn on a further item. The simplest depth sorting method is to use Cartesian y coordinate values. This works well as long as it doesn't have a sprite that occupies more than one tile space. The most efficient way of sorting depth for the isometric world is to break all the tiles into standard single tile dimensions and not allow larger images. For example, this is a tile that does not match the size of a standard tile.

To carry out the movement of the Sequencing, Overloading, Procedures, Recursive Loops and Conditionals algorithms, a slot is needed to make the character move towards the input, so in this stage arrows are used in 4 forms, namely straight, rotate left, right and jump. Where 1 input will move or move the character 1 tile from the previous tile.

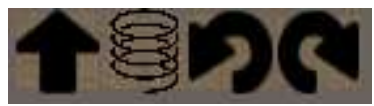

Figure 3.6. Arrow direction to make player move

For movement looping and procedural and conditional icons are used as below.

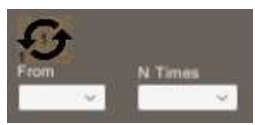

Figure 3.7. Looping and procedural icon

To understand the sorting depth correctly between characters with tiles, do the step that whenever the x and y coordinates of the character are less than the tree, the tree overlaps the character. Every time the x and y coordinates of the character are larger than the tree, the characters overlap with the tree. When a character has the same x coordinate as tiles, it is decided based on the y coordinate only, where the higher y coordinate overlaps with the other. When they have the same y coordinate, we decide based on the x coordinate only, where which has a higher x coordinate overlaps with the others.

To simplify it is necessary to draw levels sequentially starting from the farthest tiles, i.e. tile [0] [0], then draw all the tiles in each row one by one. If a character occupies a tile, we draw a ground tile first and then make a character tile. This step is applied, because the character will not be able to occupy the tile wall.

Depth sorting must be done every time a tile position changes. For example, we need to check each moving character, then update the order of depth displayed, to reflect changes in depth.

## IV. RESULT

To find out the results and helpfulness of the learning algorithm application prototype the algorithm was tested among Multimedia Nusantara University students by using the User Acceptance Test, which consisted of the following questions:

1. Is the display of this learning algorithm game app interesting?
2. Do you use this application to help you understand the use of sequence, looping and conditionals?
3. Can you use this application to help you understand the use of loops?
4. Does using this application help you understand the use of conditionals?
5. Is the presentation of this game application easy to use?

| No | Question | Percentage | | | | |
|---|---|---|---|---|---|---|
| | | Very Agree | Agree | Neutral | Not Agree | Very Not Agree |
| 1 | Is the display of this learning algorithm game app interesting? | 40% | 47% | 13% | 0% | 0% |
| 2 | Do you use this application to help you understand the use of sequence, looping and conditionals? | 13% | 87% | 0% | 0% | 0% |
| 3 | Can you use this application to help you understand the use of loops? | 27% | 73% | 0% | 0% | 0% |
| 4 | Does using this application help you understand the use of conditionals? | 13% | 80% | 7% | 0% | 0% |
| 5 | Is the presentation of this game application easy to use? | 23% | 77% | 0% | 0% | 0% |

Figure 4.1. Table User Acceptance Test Game Application Learning Algorithm

| No | Question | Grade | | | | | Count | Analisis (Jumlah) | Persentase (Analisis/5) |
|---|---|---|---|---|---|---|---|---|---|
| | | 5 | 4 | 3 | 2 | 1 | | | |
| 1 | Is the display of this learning algorithm game app interesting? | 60 | 56 | 12 | 0 | 0 | 128 | 21,33 | 4,27 |
| 2 | Do you use this application to help you understand the use of sequence, looping and conditionals? | 20 | 104 | 0 | 0 | 0 | 124 | 20,67 | 4,13 |
| 3 | Can you use this application to help you understand the use of loops? | 40 | 88 | 0 | 0 | 0 | 128 | 21,33 | 4,27 |
| 4 | Does using this application help you understand the use of conditionals? | 20 | 96 | 6 | 0 | 0 | 122 | 20,33 | 4,07 |
| 5 | Is the presentation of this game application easy to use? | 35 | 92 | 0 | 0 | 0 | 127 | 21,17 | 4,23 |
| | Average | | | | | | | 20,97 | |
| | Question Count | | | | | | | 5,00 | |
| | Result | | | | | | | 4,19 | |
| | Final Result | | | | | | | 83,87 | |

Figure 4.2. Table User Acceptance Test Result Game Application Learning Algorithm

From UAT testing taken from 30 Multimedia Nusantara University students, it was found that the average percentage of students agreed that this algorithm learning game application was interesting and helped in understanding programming algorithms was 83.87% of students.

## V. CONCLUSION

It can be concluded that, the development of algorithm learning game applications made using the near isometric projection method must pay attention to the following things in order to run well, among others, note the size of each tile, tile that will be used in the construction of this learning algorithm 26.56 degree, another thing that must be considered is how to adjust the position and level of depth of tiles between one another. In addition, character must be considered in this isometric method, where characters must be made in 8 directions for each move. Proven by using the game isometric method.

By using the User Acceptance Test (UAT) method with a sample of 30 students, it has been proven that 83.87% of students agree, it is easier to understand algorithmic learning by using the algorithmic learning game application with the near isometric projection method. It can be concluded that the application of learning algorithms is usefull for students who want to learn the basic programming and algorithms.

## VI. FUTURE WORK

Adding more algorithms level with better level design and better character control also make adding puzzle for some level, and adding juicy component, such as shaking, explode, bouncing or sound effect in game to make the game more interesting and challenging.

## VII. REFRENCES